\documentclass{ptapap}
\usepackage{amsmath}
\usepackage{amssymb}

\author{Enrique Gazta\~naga}[ICE,IEEC]
\affil[ICE]{Institute of Space Sciences (ICE, CSIC), 08193 Barcelona}
\affil[IEEC]{Institut d\'~Estudis Espacials de Catalunya (IEEC), 08034 Barcelona}
\usepackage{graphicx}	
\usepackage{amsmath}	
\usepackage{amssymb}	

\usepackage{graphicx}
\usepackage{amssymb}
\usepackage{dcolumn}
\usepackage{bm}

\usepackage{lineno}

\def\ba{\begin{array}}
\def\ea{\end{array}} 
\def\bea{\begin{eqnarray}}
\def\eea{\end{eqnarray}}
\def\beq{\begin{equation}}
\def\eeq{\end{equation}}
\def\ben{\begin{enumerate}}
\def\een{\end{enumerate}}

\def\brr{\begin{array}}
\def\err{\end{array}}

\def\calC{\S}

\title{Homogeneity and the causal boundary}

\begin{document}

\maketitle

\begin{abstract}

A Universe with finite age also has a finite causal scale $\chi_\calC$, so the metric can not be homogeneous for $\chi>\chi_\calC$, as it is usually assumed.
To account for this, we propose a new causal boundary condition, that can be fulfil by fixing  the cosmological constant $\Lambda$ (a free parameter for gravity). The resulting Universe is inhomogeneous, with possible variation of cosmological parameters on scales  $\chi \simeq \chi_\calC$. The size of $\chi_\calC$ depends on the details of inflation, but regardless of its size, the boundary condition forces $\Lambda/8\pi G $ to cancel the contribution of a  constant vacuum energy $\rho_{vac}$ to the measured $\rho_\Lambda \equiv \Lambda/8\pi G + \rho_{vac}$.   To reproduce the observed $\rho_{\Lambda} \simeq 2 \rho_m$ today with $\chi_\calC \rightarrow \infty$  we then need a universe  filled with evolving dark energy (DE) with pressure $p_{DE}> - \rho_{DE}$ and  a fine tuned value of  $\rho_{DE} \simeq 2 \rho_m$ today. This seems very odd,  but there is another solution to this puzzle. We can have a finite value of $\chi_\calC \simeq 3 c/H_0$ without the need of DE.  This scale corresponds to  half the sky at $z \sim 1$ and 60deg at $z \sim 1000$, which is consistent with  the anomalous lack of correlations  observed in the CMB.

\end{abstract}

\section{Introduction}

\label{S:1}
One of the most striking changes to Newton's  gravity proposed by Einstein is that energy gravitates. 
Scientists have since been wondering if vacuum energy $\rho_{\rm vac}$ (vacuum fluctuations, zero-point fluctuations, quantum vacuum, dark energy or aether) could also gravitate. 
 Measurements of cosmic acceleration (see e.g. \citealt{Planck2018,des2018,Tutusaus2017})  point to a model with $\Lambda$, that we refer to as $\Lambda$CDM.
Even while the accuracy and precision of  measurements have greatly improved in the last years,
the  mean values of cosmological parameters have remained similar for well over a decade (see e.g. \citealt{Gaztanaga2009,Gaztanaga2006}). 
The Friedmann-Lemaitre-Robertson-Walker (FLRW) flat metric in comoving coordinates $(t,\chi)$:
\beq
ds^2= g_{\mu\nu} dx^\mu dx^\nu = dt^2 - a(t)^2( d\chi^2 + \chi^2 d\Omega^2 )
\label{eq:frw}
\eeq
is the exact general solution for an homogeneous and isotropic flat Universe.
The scale factor, $a(t)$, describes the expansion of the Universe as a function of time. We can relate $a(t)$ to the energy content of the Universe for a perfect fluid by solving Eq.\ref{eq:rmunu}-\ref{eq:Tmunu}:
\bea
R_{00}= - \left({3 \ddot a\over{a}}\right) &=&  4\pi G (\rho + 3p)- \Lambda
\label{eq:Hubble0}
\\ \label{eq:Hubble}
H^2 &\equiv& \left({\dot{a}\over{a}}\right)^2 
=  {8\pi G\over{3}} \rho +  {\Lambda\over{3}}
\eea
where $\rho = \rho_m a^{-3}+\rho_r a^{-4} +\rho_{\rm vac}$
and $\rho_m$ is the pressureless matter density today ($a=1$),
$\rho_r$ corresponds to radiation (with pressure $p_r=\rho_r /3$)  and $\rho_{\rm vac}$  represents vacuum energy ($p_{vac}= -\rho_{\rm vac} $). 
 One can argue that 
$\Lambda$ is indistinguishable from $\rho_{vac}$, because
equations are degenerate to the combination:
\beq
\rho_{\Lambda}  \equiv  \rho_{\rm vac} + \frac{\Lambda}{8\pi G}  .
\label{eq:rhoHlambda}
\eeq
Here we take $\Lambda$ to be a fundamental constant, while $\rho_{vac}$ depends on the energy content. The measured $\rho_\Lambda$ is extremely small compared to what we expect for $\rho_{\rm vac}$. 
Moreover, $\rho_\Lambda \simeq 2.3\rho_m$ today, which is a remarkable coincidence. 
Possible solutions to this puzzle are:  I) $\Lambda=0$, so that
 $\rho_\Lambda$ originates only from $\rho_{\rm vac}$ or dark energy (DE) \citep{Weinberg1989,Elizalde1990, Carroll,Huterer,Elizalde2006}, II)  $\rho_{\rm vac}=0$ and we need $\Lambda$ or Modified Gravity \citep{Lobo2001,Gaztanaga2002,Lue,Nojiri2017} or III) there is a cancellation between $\Lambda$ and $\rho_{\rm vac}$,  as will be shown here.

Eq.\ref{eq:frw}-\ref{eq:Hubble} represent a mathematical extrapolation. A physical explanation requires a mechanism to produce homogeneity.
Regardless of the details of inflation or the early Universe,
a Universe of finite age will only be causally homogeneous 
for scales smaller than some cut-off $\chi<\chi_\calC$.
We need some boundary condition at $\chi=\chi_\calC$ to account for the lack of causality (and therefore homogeneity) at larger scales.  This boundary results in cosmic acceleration and a cancellation between $\Lambda$ and $\rho_{vac}$.
In \S2  we view this problem in Classical Physics, while in \S3 we present the relativistic version. In \S4 we estimate the size of the causal Universe and discuss the implications for inflation and CMB. We end with some Discussion and Conclusions.

\section{Hooke's law}

A key property of Gravity  in Classical Mechanics is Gauss law.
The acceleration $\vec{\mathrm{g}}$ created by a point mass (or charge)  at distance $\vec{\mathrm{r}}$  is such
that a spherical shell of arbitrary radial density $\rho(r)$
produces a field which is identical to a point source of equal mass (or charge) $m$ in its center. 
The solution \citep{Wilkins} is more general than Newton's law:
\bea
\vec{\mathrm{g}} \equiv  -\vec{\nabla} \phi &=& -\left( \frac{G m}{r^3} - \frac{\Lambda}{3} \right) \vec{r}
\nonumber \\
\nabla^2 \phi &=& 4\pi G \rho_m -\Lambda .
\label{eq:hooke}
\eea
Using Stokes Theorem, the field  produces a flux around 2D closed surface $\partial V$:
\beq
\Phi= 
\oint_{\partial V} ~d\vec{\mathrm{r}} ~ \vec{\mathrm{g}}  = 
- \int_V  dV \left( 4\pi G  ~\rho_m  - \Lambda  \right) 
\label{eq:flux00}
\eeq
so the flux only depends on the total mass inside the boundary $\partial V$.
Note the second term with $\Lambda$ in  Eq.\ref{eq:hooke}-\ref{eq:flux00} which corresponds to Hooke's law, i.e. proportional to distance. These of course are the same equations that come from General Relativity in the Newtonian limit (see below).
Thus $\Lambda$ in Eq.\ref{eq:hooke}-\ref{eq:flux00} is allowed by the symmetries of gravity. In fact Newton, and other scientist, had already noticed this \citep{CalderLahav2008}. 

Physicist assume that particles should be free at infinity, because of lack of causality. This is  why boundary terms are usually neglected at infinity. In the same spirit,
 we will require here that test  particles should be free  ($\vec{\mathrm{g}}=0$) or more relevant for a fluid: that boundary terms should be zero ($\Phi =0$), when outside causal contact, $r> r_\calC$.
This condition for $ r_\calC \Rightarrow \infty$ 
requires $\Lambda \Rightarrow0$, as otherwise 
$\vec{\mathrm{g}}$ and $\Phi$ diverge.
Observational evidence that $\Lambda \neq 0$ may then indicate that $r_\calC$ is finite. 
This agrees with the finite age of the Universe, which also implies that
 $r_\calC < \infty$.  
 From Eq.\ref{eq:flux00} the boundary condition $\Phi(r>r_\calC)=0$, implies:
 \beq 
  \Lambda= 4\pi G \rho_m(r<r_\calC)
  \eeq 
  which  can provide an explanation for the coincidence $\rho_\Lambda \sim 2 \rho_m$.
  Lets next explore this same argument in General Relativity. For this we first need to see what is the relativistic version of Eq.\ref{eq:flux00}.

\section{Relativistic case}
\label{sec:infinite}

The symmetries of Einstein's  field equations allow for a cosmological constant $\Lambda$ term (\citealt{Landau1971,Weinberg1972}):

\beq
R_\mu^\nu + \Lambda \delta_\mu^\nu = 8\pi~G~(T_\mu^\nu-\frac{1}{2} \delta_\mu^\nu T) ,
\label{eq:rmunu}
\eeq
For a  perfect  fluid with  density $\bar{\rho}$ and pressure  $\bar{p} \equiv \omega \bar{\rho}$:

\beq
T_\mu^\nu =  (\bar{p}+\bar{\rho}) u_\mu u^\nu - \bar{p} \delta_\mu^\nu  
\label{eq:Tmunu}
\eeq
where both $\bar{p}$ and $\bar{\rho}$ could change with space-time.

\subsection{The generalised Gauss's law}

Consider perturbations around 
Minkowski metric $\eta_{\mu\nu} $ (i.e. around empty space):

\beq 
g_{\mu\nu} = \eta_{\mu\nu} + h_{\mu\nu}
\label{eq:h}
\eeq 
 where $h_{\mu\nu}$ are small corrections. 
 To linear order in  $h_{\mu\nu}$ 
 we have $R_{00} = 1/2 ~ \square h_{00}$ \citep{Landau1971},
so that:

\beq
R_{00}  = R_0^0 =  \nabla_\mu \nabla^\mu \phi
\label{eq:R00}
\eeq
where $\phi \equiv  h_{00}/2$ is the gravitational potential.
We can now use the right side o Eq.\ref{eq:rmunu}  to calculate $R_0^0$ by contracting  
$T_\mu^\mu$ in Eq.\ref{eq:Tmunu}. We consider observers moving with the fluid so that the 3D velocity $u^i=0$  so that  $u^\mu u_\mu = u^0 u_0=1$ :

\beq
R_0^0
= 
 4\pi G  ~(\bar{\rho}+ 3 \bar{p})  - \Lambda ,
\eeq
where $\bar{\rho}$ and $\bar{p}$ are perturbations around empty space. Combined with Eq.\ref{eq:R00},  this results in:
\beq
\nabla_\mu \nabla^\mu \phi = 4\pi G  ~(\bar{\rho}+ 3 \bar{p})  - \Lambda
\label{eq:R00b}
\eeq
which is the relativistic generalization of Poisson equation Eq.\ref{eq:hooke}.
Thus the relativistic version of Gauss law in Eq.\ref{eq:flux00} is then 
\beq
\Phi =  - \int_{M}  \sqrt{-g} ~~d^4x  
~\left[  4\pi G  ~(\bar{\rho}+ 3 \bar{p})  - \Lambda \right]
\label{poisson3}
\eeq
where $M$ is the 4D volume inside the 3D surface $\partial M$. 

\subsection{Causal Boundary condition}
\label{sec:boundary}

As discussed in the introduction, 
causality can only be efficient for $\chi<\chi_\calC$.
Larger scales can have no effect on the metric, which mathematically is equivalent to $T_{\mu\nu}$
equal to zero for $\chi> \chi_\calC$. In Eq.\ref{eq:Tmunu}, this corresponds to:
\beq
\bar{\rho}=\rho ~{\cal{H}}(\chi_\calC-\chi) 
~;~ \bar{p}=p ~{\cal{H}}(\chi_\calC-\chi)
\label{eq:Heaviside}
\eeq
where ${\cal{H}}$ is the Heaviside step function (or a smoothed version of it) and $\rho$ and $p$ are quantities inside $\chi_\calC$. 
This requires a non-homogeneous solution for the metric of the Universe
(see \citealt{Gaztanaga2019}).
But it does not mean that space is physically empty for $\chi>\chi_\calC$: it just can't have any effect on the metric of our local observer.
An observer situated at the edge of our causal boundary will find a similar solution, but could measure different cosmological parameters, because she sees a different patch of the initial conditions.  There should be a 
 smooth background across disconnected regions with  an infrared cutoff in the spectrum of inhomogeneities for $\chi>\chi_\calC$.
Solutions in different regions could be matched as in  \cite{Sanghai-Clifton}.

On scales $\chi<\chi_\calC$ we  have a homogeneous expanding Universe with $\bar{\rho}=\rho$. On larger scales we want particles to be free, i.e. to approach Minkowski metric $\eta_{\mu\nu}$ as in Eq.\ref{eq:h}.
Thus we require the boundary term $\Phi(\chi>\chi_\calC)=0$ in Eq.\ref{poisson3}. This implies: 

\beq
\frac{\Lambda}{8\pi G} =   \frac{1}{2M_{\calC}} \int_{M_{\calC}}   \sqrt{-g} d^4x  ~  (\rho+ 3p) ,
\label{eq:rhoH0}
\eeq
where $M_{\calC}$ is the volume inside the lightcone to the surface $\partial M_{\calC}$, where  $\chi=\chi_\calC$.
Recall how Eq.\ref{poisson3} was obtained in the weak field limit.
The exact result for a general (non-homogeneous) spherically symmetric metric is  \citep{Gaztanaga2019}:
\beq
\frac{\Lambda}{8\pi G} =   \frac{ \int_{M_{\calC}}   \sqrt{-g} d^4x  ~  (\rho+ 3p) (1 - H^2 r^2)}
{2\int_{M_{\calC}}   \sqrt{-g} d^4x ~ (1 - H^2 r^2)   } ,
\label{eq:rhoH}
\eeq
the additional factor $(1-H^2 r^2)$ accounts for deviations from the weak field expansion used in Eq.\ref{poisson3}. 
We shall check below  how well this weak field approximation works.

\subsection{Vacuum Energy does not gravitate}
\label{sec:vac}
Inside $\chi<\chi_\calC$, we can use  Eq.\ref{eq:frw}-\ref{eq:Hubble} 
with $\rho= \rho_m+\rho_r+\rho_{\rm vac}$ and $p= \rho_r/3-\rho_{\rm vac}$, so that 
we can write Eq.\ref{eq:rhoH}  as:
\beq
\frac{\Lambda}{8\pi G} = \frac{\rho_m(\calC)}{2} + \rho_r(\calC) - \rho_{\rm vac}
\equiv \rho_{\calC} - \rho_{\rm vac} ,
\label{eq:m2}
\eeq
where $\rho_{\calC}$ is the matter and radiation contribution in the integral of Eq.\ref{eq:rhoH}. The values of $\rho_m$ and $\rho_r$ evolve with space-time, so that $\rho_{\calC}$ is the average contribution inside the volume $M_\calC$, while the vacuum density contribution is constant (by definition).  We can combine Eq.\ref{eq:rhoHlambda} with  Eq.\ref{eq:m2}:

\beq
\rho_{\Lambda}  \equiv \frac{\Lambda}{8\pi G}  + \rho_{\rm vac} =  \rho_{\calC} - \rho_{\rm vac} +   \rho_{\rm vac} = {\rho}_{\calC} ,
\label{eq:rhoH2}
\eeq
which shows that vacuum energy cancels out and can not change the observed value of $\rho_{\Lambda}$.

\subsection{Effective Dark Energy (DE)}
\label{sec:DE}

If  vacuum energy suffers a phase transition or changes in some other way, as  is believed to have happened during inflation, then this cancellation will not necessarily happen and an evolving  $\rho_{\rm vac}$ (which we usually call Dark Energy) could contribute to the effective value of $\rho_{\Lambda}$. Consider the more general case of DE:
\bea
\rho_{DE}(a)  &=& \rho_{\rm vac} + \rho_{DE} ~a^{-3(1+\omega)}
\label{eq:DE}
\\ \nonumber
p_{DE}(a)  &=& -\rho_{\rm vac} + \omega ~\rho_{DE} ~ a^{-3(1+\omega)} ,
\eea
where only one component of DE is evolving.
We  then have from Eq.\ref{eq:rhoH} and Eq.\ref{eq:rhoHlambda}:
\beq
\rho_{\Lambda} 
= {\rho}_{\calC} +
 \rho_{DE} [1 +  \frac{1+3\omega}{2} \hat{a}_{\calC}^{-3(1+\omega)}] .
 \label{eq:rhoH3b}
\eeq
where $\hat{a}_{\calC}$ is some mean value of $a$ in the past light-cone of $a_\calC$ in Eq.\ref{eq:rhoH}.
This reduces to  $\rho_\Lambda = \rho_\calC$ for $\omega=-1$.
For $a_\calC \Rightarrow \infty$ 
we have $\rho_\calC \Rightarrow 0$  because  $\rho_m(a)$ and $\rho_r(a)$ tend to zero as we increase $a_\calC$. The same happens with $\hat{a}_{\calC}^{-3(1+\omega)}$ for $\omega >-1$, so that:
\beq
\rho_\Lambda  = \rho_{DE}   ~~ \rm{for}~~ {a}_\calC \Rightarrow \infty  ~~\rm{\&} ~~ \omega>-1 .
\label{eq:rhoDE}
\eeq
So evolving DE  could produce the observed cosmic acceleration in an infinitely large Universe. 
This solution does not explain why $\rho_{\Lambda} = \rho_{DE} \simeq 2\rho_m$. The original motivation to introduce DE was to explain how vacuum energy $\rho_{\rm vac}$ could be as small as the measured $\rho_\Lambda$ \citep{Weinberg1989,Huterer}. But we have shown in Section \ref{sec:vac} that the causal boundary condition explains why $\rho_{\rm vac}$ does not contribute to $\rho_\Lambda$ and also results in $\rho_{\Lambda} \simeq 2\rho_m$. This removes the motivation to have DE, as it represents an unnecessary complication of the model.

\section{The size of the Causal Universe}
\label{sec:size}

We  assume in this section that vacuum energy is constant after inflation ($\omega=-1$). In this case Eq.\ref{eq:rhoH}-\ref{eq:rhoH2} give:
\beq
\rho_{\Lambda} = \rho_{\calC} =
 \frac{\int_{M_{\calC}}  \sqrt{-g} d^4x  (\rho_m + 2\rho_r) ~ (1 - H^2 r^2)}
 {\int_{M_{\calC}}  \sqrt{-g} d^4x ~ (1 - H^2 r^2)}.
\label{eq:rhoH4}
\eeq

The horizon after inflation
(see Eq.6.20 in \citealt{Dodelson} and Eq.\ref{eq:eta} below) 
is:
\beq
\chi(a) = \eta(a) - \eta(a_e)
\label{eq:chia}
\eeq
where $a_e$ represents the end of inflation. We then have $\chi_\calC = \chi(a_\calC) = \eta(a_i)$
where $a_{\calC}$ is the time when the causal boundary enters the horizon after inflation and $a_i$ the begining of inflation.  Fig.\ref{Fig:horizon} illustrate this.
We calculate $\rho_{\calC}$  in Eq.\ref{eq:rhoH4} as the integral to $\chi_{\calC}$ in the light-cone:

\beq
\rho_{\calC}  = 
~ \frac{\int_{0}^{\chi_{\calC}}  d\chi ~ \chi^2 ~{a^3}~(\rho_m a^{-3} + 2\rho_r a^{-4})~ (1 - a^2 \chi^2 H^2 )}
{2 \int_{0}^{\chi_{\calC}}  d\chi  ~\chi^2 ~a^3 ~ (1 -  a^2 \chi^2 H^2)} ,
\label{Eq:rhoHchi}
\eeq
where $a=a(\chi)$ in Eq.\ref{eq:chia}.
 For $H(a)$ we use Eq.\ref{eq:Hubble}
with $\Omega \equiv \rho/\rho_c$ 
$\rho_c=  3H_0^2 / 8\pi G$, 
$\Omega_r=4.2 \times 10^{-5}$ \citep{Planck2018}
and flat Universe
$\Omega_m=1-\Omega_\Lambda-\Omega_r$. 
We can use Eq.\ref{Eq:rhoHchi} to solve $\Omega_\Lambda=\Omega_{\calC}$ numerically 
given $\Omega_\Lambda=0.69 \pm 0.01$ \citep{Planck2018}. We find:

\bea
\chi_{\calC} &=& \left( 3.034 \pm 0.004 \right) \frac{c}{H_0} 
\label{eq:chi_H} \\
a_{\calC} &=& 0.85 \pm 0.003 .
\label{eq:a_H}
\eea
to be compared to $a_0=1$  and $\chi_0=3.200{c\over{H_0}}$ today. 
Because $\chi_\calC$  is smaller than $2\pi$ times our observable horizon, 
we should be able to see this horizon in our past lightcone
at  $\theta_\calC=\chi_\calC/\chi$. At $z\simeq 1$  about half of the sky is causally disconnected. 
At larger redshifts this boundary tends to a fix value $\sim 60$ deg. depending on  $\chi_\calC$ (and therefore $\Lambda$).
This has implications for CMB observations (see  section \ref{sec:CMB}).

 If we set $\Omega_r=0$ we find $a_{\calC}=0.86$ and  $\chi_{\calC}=3.081$, so our results are not very sensitive to the details of the Early Universe. 
If we neglect the $(1-H^2r^2)$ factor in Eq.\ref{eq:rhoH} (which accounts from deviations from the weak field limit) we find instead:
$\chi_{\calC} =  3.138$ and $a_{\calC} = 0.94$, which indicates small but significant deviations from  the 
weak field approximation in Eq.\ref{eq:rhoH0}. 


\subsection{Inflation and the coincidence problem}
\label{sec:coincidence}

\begin{figure}
\centering \includegraphics[width=0.7\textwidth]{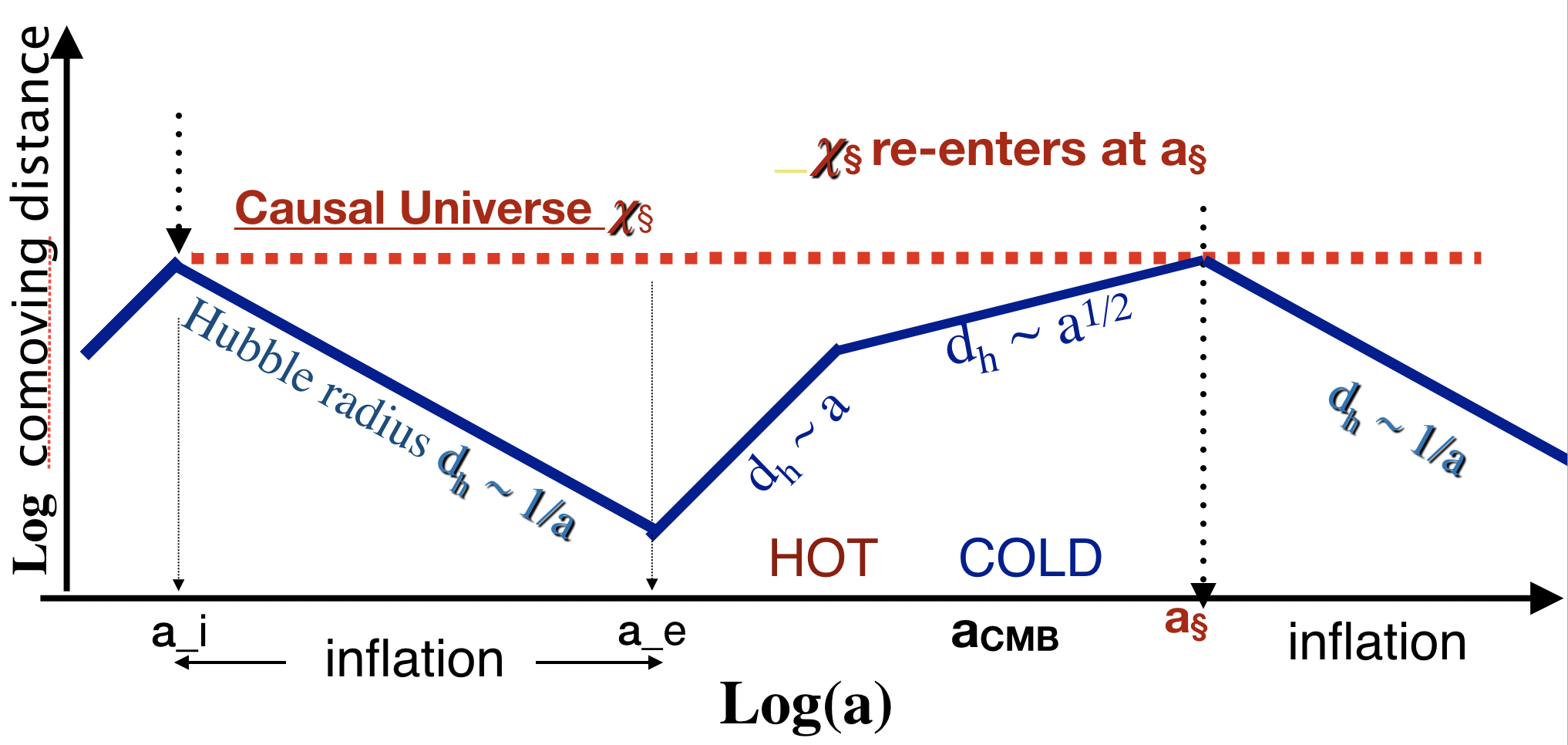}
\caption{Comoving 
Hubble radius  $d_H = c/(aH) $ (blue line) as a function of the scale factor $a$.  The Causal Universe $\chi_\S$ is identified with the region inside the largest causally connected scale at the beginning of inflation (red dashed line).}
\label{Fig:horizon}
\end{figure}

Particles separated by distances larger than the comoving Hubble radius 
$d_H(t)=c/[a(t)H(t)]$ can't communicate at time $t$. Distances larger than the horizon
\beq
\eta(a) = c \int_{0}^{t}  \frac{dt}{a(t)} = c \int_{0}^{a}  {d\ln(a)} ~d_H(a) ,
\label{eq:eta}
\eeq
have never communicated.
We know from the cosmic microwave background (CMB) and large scale structure (LSS) that the Universe was very homogeneous on scales that were not causally connected (without inflation). 
This either means that the initial conditions where acausally  smooth to start with or that there is a mechanism like inflation \citep{Dodelson,Liddle1999,Brandenberger} which inflates homogeneous and causally connected regions outside the Hubble radius.
 This  allows the full observable Universe to originate from a very small causally connected  homogeneous patch, which here we call the Causal Universe,  $\chi_\calC$. 
During inflation, $d_H$  decreases  which freezes out communication on comoving scales larger than the horizon  $\chi_\calC \simeq \eta(a_i)=d_H(t_i)$ when inflation begins, at $a_i=a(t_i)$. 
When inflation ends, radiation from reheating makes $d_H$ grow again. 
Eq.\ref{eq:rhoH4} indicates that when the causal boundary re-enters the Horizon the expansion becomes dominated by $\rho_\Lambda$. This is because $\rho_m(a_\calC)<\rho_\calC=\rho_\Lambda$, as density decreases with the expansion. This results in another inflationary epoch at $a=a_\calC$ which keeps the Causal Universe frozen (see Fig.\ref{Fig:horizon}).  Thus, causality  can only play a role for comoving scales $\chi<\chi_\calC$.  The Causal Universe $\chi_\calC$ is therefore fixed before inflation in comoving coordinates and is the same for all times, while the horizon $\eta$ and $d_H$ change with time.

 We can now recast the coincidence problem (why $\rho_{\Lambda} \simeq 2.3 \rho_m$?)
into a new question: why we live at a time which is close to  $a_\calC$? 
In terms of anthropic reasoning  \citep{Weinberg1989,2003PhRvD..67d3503G}, 
at earlier times the Universe is dominated by radiation and there are no stars or galaxies to host observers. 
Closer to $a\simeq a_\calC$ the Universe is dominated by matter and there are galaxies and stars with planets and potential observers.
At later times $\Omega_\Lambda \simeq 1$ and galaxies will be torn apart by the new inflation. 
Moreover, $a_\calC$ has the largest Hubble radius (see Fig.\ref{Fig:horizon}) with the highest chances to host observers like us.
There is nothing too special about this coincidence.
Ultimately, the reason why $\chi_\calC \sim 3c/H_0$  reside in the details of inflation: when inflation begins $a_i$ and ends $a_e$ (see Fig.\ref{Fig:horizon}).
This recasts the coincidence problem into an opportunity to better understand inflation and the origin of homogeneity. We propose here to identify $\chi_\calC=\eta(a_i)$ with the comoving  horizon before inflation begins at time $t_i$, $H_i=H(t_i)$ or $a_i=a(t_i)$:

\begin{equation}
a_i H_i =  c\chi_{\calC}^{-1} \simeq \left( 0.328 \pm 0.004\right) H_0
\label{eq:aH}
\end{equation}
The Hubble rate during inflation $H_I$ is proportional to the energy of inflation. During reheating this energy is converted into radiation:
$H_I^2 \simeq \Omega_r ~H_0^2 ~ a_e^{-4}$, with
$a_e \equiv e^{N} a_i$. We can combine  with Eq.\ref{eq:aH} to find:

\beq
a_i \chi_{\calC} = \frac{H_i}{H_I} e^{-2N} ~ \Omega_r^{1/2} ~ (\chi_{\calC}^2 H_0/c) \simeq 4 \times 10^8 l_{\rm Planck} 
\label{eq:achi}
\eeq
where for the second equality we have used the canonical value of $N \simeq 60$
and $H_i \simeq H_I$, which also yields $a_i \simeq 1.56 \times 10^{-53}$
and $H_I \sim 10^{10}$ GeV.
 The condition  $a_i \chi_{\calC}> l_{\rm Planck}$ requires $N<70$, close to the value found in \cite{DodelsonHui2003}.

\subsection{Implications for CMB}
\label{sec:CMB}

\begin{figure}
\includegraphics[width=.7\linewidth]{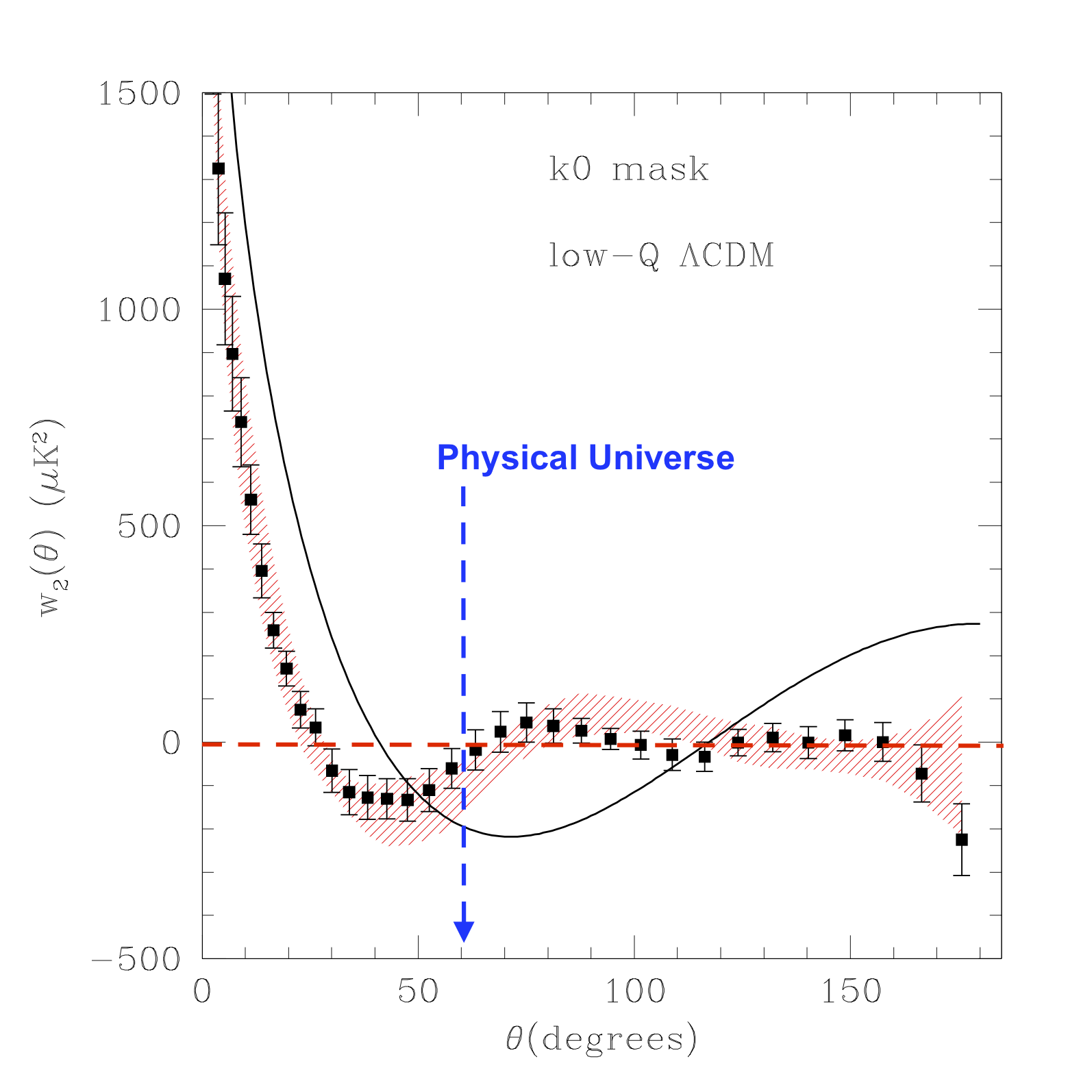}
\caption{
Two-point correlation function of measured CMB temperature fluctuations in WMAP (points with errorbars) as a function of angular separation. The black continuous line shows the  $\Lambda$CDM prediction for an infinite Universe. 
}
\label{Fig:CMB}
\end{figure}

The (look-back) comoving distance to the surface of last scattering $a_* \simeq 9.2 \times 10^{-4}$ \citep{Planck2018} is
$\chi_{CMB} = \eta(1)- \eta(a_*) \simeq 3.145 ~ \frac{c}{H_0}$. This is just slightly larger than our estimate for $\chi_\calC$ in Eq.\ref{eq:chi_H}.
Thus, we would expect to see no correlations in the CMB on angular scales  $\theta > \theta_\calC \equiv \frac{\chi_\calC}{\chi_{CMB}} \simeq 56-62$ degrees 
for $\Omega_\Lambda =\Omega_\calC = 0.7-0.5$.
The lack of structure seen in the CMB on these large scales is one of the well known anomalies in the CMB data, see \cite{Schwarz2016} and references therein. 
Fig.\ref{Fig:CMB} (from \citealt{Gaztanaga2003}) shows a comparison of the measured CMB temperature correlations (points with error-bars) with the $\Lambda$CDM prediction  for an infinite Universe (continuous line). There is a very clear discrepancy.
If we suppress the large scale modes (multipoles $l<5$)
in the $\Lambda$CDM simulations, the agreement is much better (shaded red area in Fig.\ref{Fig:CMB}). 

We can also predict $\Omega_\Lambda$ from the lack of CMB correlations. From Fig.\ref{Fig:CMB} we roughly estimate $\theta_\calC \simeq 60 \pm 2$ deg. to find (using Eq.\ref{Eq:rhoHchi}) $\Omega_\Lambda = 0.6 \pm 0.1$. 
But note that this rough estimate does not take into account the foreground (late) ISW and lensing effects \citep{Fosalba,ISW}, which will typically  reduce $\theta_\calC$ because they add non primordial correlations to the largest scales.
This requires further investigation. Also note  that this estimate for $\Omega_\Lambda$
corresponds to the size of disconnected regions at the location of the CMB, which might be slightly different to the value near us, as we see a different patch of the primordial Universe (see bellow).
Note also that there are temperature differences on scales larger $\theta_\calC$, but they are not correlated, as expected in causality disconnected regions.
Nearby regions are connected which creates a smooth transition across disconnected regions.

\section{Discussion and Conclusions}

$\Lambda$CDM in Eq.\ref{eq:frw}-\ref{eq:Hubble}  assumes that $\rho$ is constant everywhere at a fixed comoving time.
This requires acausal initial conditions \citep{Brandenberger} unless there is inflation, where a tiny homogeneous and causally connected patch, the Causal Universe $\chi_\calC$, was inflated to be very large today. Regions larger than  $\chi_\calC$ are out of causal contact. Here we require that 
test particles become free (or the relativistic flux is zero)
as we approach $\chi_\calC$.
This leads to Eq.\ref{eq:rhoH}, which is the main result in this paper.
If we ignore  the vacuum, this condition requires:
$\Lambda  = 8\pi G \rho_{\calC}$, where $\rho_\calC$ is the  matter and radiation inside 
$\chi_\calC$ (Eq.\ref{eq:rhoH4}). For an infinite Universe
($\chi_{\calC} \rightarrow \infty$) we have $\rho_{\calC} \Rightarrow 0$ which requires $\Lambda \Rightarrow 0$. This is also what we find in classical gravity with a $\Lambda$ term, because Hooke's term diverges at infinity (see Eq.\ref{eq:hooke}).  So  the  fact that $\Lambda \neq 0$   could indicate that  $\chi_\calC$ is not  infinite. Adding vacuum  $\rho_{\rm vac}$ does not change this argument because  $\rho_\Lambda \equiv \Lambda/ 8\pi G + \rho_{\rm vac} = \rho_{\calC}$ turns out to be independent of $\rho_{\rm vac}$
(see Eq.\ref{eq:rhoH2}). 
Thus, whether the causal size of the Universe is finite or not, $\rho_{\rm vac}$ can not gravitate! The cancellation between $\Lambda$ and $\rho_{\rm vac}$ is a direct consequence of the boundary condition and it also implies that deSitter Universe (empty with a cosmological constant) is not causal (as it produces curvature being empty).
In classical terms of Eq.\ref{eq:hooke}, it will produce a  divergent gravitational force.

For constant vacuum ($\omega \equiv p/\rho =-1$), we find  $\chi_\calC \simeq 3 c/H_0$ 
for  $\Omega_\calC= \Omega_\Lambda \simeq 0.7$. 
We can also estimate $\chi_\calC$ as $c/(a_i H_i)$ 
when inflation begins, see Eq.\ref{eq:aH}. After inflation  $\chi_\calC$ freezes out until it re-enters causality at $a_\calC \simeq 0.85$, close to now ($a=1$). This starts a new inflation (as $\rho_\Lambda = \rho_{\calC}> \rho_m$) which keeps the causal boundary frozen. 
Thus a finite  $\chi_\calC$  explains why $\rho_\Lambda \simeq 2\rho_m$.
It also predicts that CMB temperature should not be correlated above $\theta> \theta_\calC \simeq 60$ deg. A prediction that matches observations
(see Fig.\ref{Fig:CMB}). One can  reverse this argument to use the lack of CMB correlations above
$\theta_\calC \simeq 60$ deg, to estimate $\chi_\calC \simeq \theta_\calC \chi_{CMB}$. Together with condition $\rho_\calC=\rho_\Lambda$, this  provides a  prediction of $\Omega_\Lambda \simeq 0.6 \pm 0.1$, which is independent of other measurements for $\Omega_\Lambda$.
 This is slightly lower than other  local measurements, but more work is needed to account for the late ISW and lensing  
 and to interpret the CMB measurements with a metric that is not homogeneous \citep{Gaztanaga2019}.

Note that because the Universe is not strictly homogeneous outside a causal region, the causal boundary for observers far away from us could be slightly different from ours, because they see a different patch of the Universe which could have slightly different energy content. Continuity across nearby disconnected regions forces these differences to be small, but it is impossible to quantify this without a model for the initial conditions and a better understanding of the process that generates the primordial homogeneity. 
In general such differences could affect structure formation, galaxy evolution and CMB observations. The fact that we can measure a concordance picture from different observations with the $\Lambda$CDM model indicates that these differences must be small. But tensions between measurements of cosmological parameters from very different redshifts (eg between CMB and local measuremnts) could be related to such in-homogeneities, rather than to evolution of the Dark Energy (DE) equation of state or other more exotic explanations.

For $\chi_\calC>> 3c/H_0$  we can not explain cosmic acceleration with $\omega=-1$, because  the resulting $\rho_\Lambda$  in Eq.\ref{eq:rhoH4} would be very small. We need evolving DE with  equation of state $\omega_{DE}>-1$ and $\rho_\Lambda = \rho_{DE}$ today.  But DE gives no clue as to why  $\rho_{DE} \simeq 2 \rho_{m}$ today and can not explain  the anomalous lack of CMB correlations at large scales. 
We apply Occam's razor to argue that there is no need for DE or Modify Gravity: measurements of cosmic acceleration and CMB can be explained by the finite age of
 our Universe using the standard Einstein's field equations and standard matter-energy content.

\acknowledgments{
I want to thank A.Alarcon, J.Barrow, C.Baugh, R.Brandenberger, G.Bernstein, M.Bruni, S.Dodelson,  E.Elizalde,  J.Frieman, M.Gatti, L.Hui, D.Huterer, A. Liddle, P.J.E. Peebles, R.Scoccimarro and S.Weinberg for their feedback.
This work has been supported by MINECO  grants AYA2015-71825, LACEGAL Marie Sklodowska-Curie grant No 734374 with ERDF funds from the EU  Horizon 2020 Programme. IEEC is partially funded by the CERCA program of the Generalitat de Catalunya. }

\bibliographystyle{ptapap}
\bibliography{gaztanaga}

\end{document}